\documentclass{pasj00}
\draft

\begin{document}

\title{Cryogenic Capacitive Transimpedance Amplifier for Astronomical Infrared Detectors}

\author{Hirohisa \textsc{Nagata}}%
\affil{Advanced Technology Center, \\
National Astronomical
Observatory of Japan, Mitaka, Tokyo, 181-8588, Japan}
\email{hirohisa.nagata@nao.ac.jp}
\author{Hiroshi \textsc{Shibai}, Takanori \textsc{Hirao}, Toyoki \textsc{Watabe},Yasunori \textsc{Hibi} and Mitsunobu \textsc{Kawada}}
\affil{Graduate School of Science, Nagoya University, Furo-cho, Chikusa-ku, Nagoya, 464-8602,Japan}
\author{Manabu \textsc{Noda}}%
\affil{Nagoya City Science Museum, Nagoya, Aichi, 460-0008,
Japan.}

\and
\author{Takao {\sc Nakagawa}}
\affil{Infrared Astrophysics Group, Institute of Space and Astronautical
Science, \\
Japan Aerospace Exploration Agency, Sagamihara, Kanagawa
229-8510, Japan}

\Received{2003 June 16}
\Accepted{2003 November 3}

%


\maketitle

\begin{abstract}
We have developed a new capacitive transimpedance amplifier (CTIA) that can be operated at 2 K, and have good performance as readout circuits of astronomical far-infrared array detectors. The circuit design of the present CTIA consists of silicon p-MOSFETs and other passive elements. The process is a standard Bi-CMOS process with 0.5 $\mu$m design rule. The open-loop gain of the CTIA is more than 300, resulting in good integration performance. The output voltage swing of the CTIA was 270 mV. The power consumption for each CTIA is less than 10 $\mu$W. The noise at the output showed a 1/f noise spectrum of 4 $\mu$V/${\sqrt {\rm Hz}}$ at 1 Hz. The performance of this CTIA nearly fulfills the requirements for the far-infrared array detectors onboard ASTRO-F, Japanese infrared astronomical satellite to be launched in 2005. 
\end{abstract}

\section{Introduction}
Establishment of cryogenic readout circuits is one of the most important technologies for astronomical application in the far-infrared wavelength region (30 - 300 $\mu$m), because preamplifier circuits must be placed in the very vicinity of the detectors at a cryogenic temperature (below 4 K) in order to suppress electromagnetic interference (EMI) and microphonic noises. Recently, larger-formatted, two-dimensional (2D) array detectors are required for the far-infrared astronomical observation, which implies the development of larger-formatted, preamplifier array operable at the cryogenic temperature.

However, at cryogenic temperatures below 4.2 K, most of semiconductor devices such as Si JFET and bipolar transistors cannot be used due to the carrier freeze-out effect. The most probable silicon device that can be used at lower than 4.2 K is the silicon metal-oxide-semiconductor field-effect transistor (MOSFET). Of course, we can use materials other than silicon, such as GaAs, Ge, etc. However, the silicon MOSFET has a great advantage that the mass production technology established for commercial usage is available for larger-formatted array circuits. For example, in the field of near-infrared astronomy, silicon MOSFETs have already been widely used as readout circuits of near-infrared detector arrays at 15 - 80 K. Additionally, silicon MOSFETs can be operated with low power dissipation, and very suitable as parts of cryogenic readout circuit. On the other hand, silicon MOSFETs frequently show anomalous current-voltage characteristics such as "kink" and "hysteresis" below liquid helium temperature [1].

  Several approaches have been dedicated to develop cryogenic readout circuits until now. The cryogenic readout circuits used for ISOPHOT [2] onboard the Infrared Space Observatory (ISO) [3] contained an amplifier with both n-MOSFETs and p-MOSFETs that was designed to eliminate anomalies mainly from n-MOSFET [4]. For MIPS (The Multiband Imaging Photometer for SIRTF) on board the Space Infrared Telescope Facility (SIRTF) [5], the cryogenic readout circuits called CRC696 are employed [6]. Special wafers that has very thin, lightly doped epitaxial layers over degenerate substrate was used for CRC696, and it solved the problem of DC instability of the readout circuit at less than 20 K [7].

On the other hand, several unique efforts have been successfully made for development of cryogenic readout circuit technology in Japan [8, 13]; p-MOSFETs of a special structure, which can be fabricated by a standard BiCMOS process, show good performances below liquid helium temperature as follows. The drain resistance and tranceconductance of the p-MOSFETs at the drain current of -1 $\mu$A and at the drain-source voltage of -1 V are $\sim$2 M$\Omega$ and $\sim$35 $\mu$S, respectively [8]. The noise voltage at 1 Hz is 1.9 $\mu {\rm V_{rms}}$/${\sqrt {\rm Hz}}$ [8]. Moreover a cryogenic amplifier consisting of this p-MOSFET has an open-loop gain of more than 1000. On the basis of the previous work, the present work is focused in developing a capacitive transimpedance amplifier (CTIA) by using this cryogenic amplifier developed by Hirao et al. [8]. Moreover, a 60-channel CTIA array with low-pass filters and multiplexers (MPXs) are also designed by the same technology for far-infrared Ge:Ga array detectors of the Far Infrared Surveyor (FIS) [9] on board ASTRO-F [10]. Therefore, by adopting this new MOSFET structure, highly developed, commercially based silicon MOS technology enables us to make a more complicated cryogenic preamplifier array. Moreover, it is also demonstrated that we will be able to make highly integrated LSI that can be operated even at 2 K by using the same technology. In the far-infrared and submillimetric wavelength regions, we will be able to make much more integrated array detectors. In section II, we describe the design and fabrication. The results of the performance measurement and discussion are presented in section III and IV, respectively.

\section{DESIGN AND FABRICATION}
\subsection{Requirements for FIS onboard ASTRO-F}
The Far-Infrared Surveyor (FIS), one of the two focal plane instruments of the ASTRO-F, covers the wavelength range of 50-200 $\mu$m. The FIS has two sets of two-dimensional Ge:Ga photoconductor array systems: the shorter wavelength band array (SW) and the longer wavelength band (LW) array. The SW is composed of two independent direct-hybrid array detectors of 3 $\times$ 20 pixels operated at 2.5 K [11,12]. The LW has a stressed Ge:Ga array detector of 5  $\times$ 15 pixels operated at 1.8 K [13]. As the readout amplifier for these arrays, the CTIA-type preamplifier was adopted because of the following reason. The Ge:Ga photoconductor is the best available detector device for the astronomical far-infrared observation at present. However, its responsivity severely depends on the bias electric field in the detector, i.e., the bias voltage externally applied. The classical integration-type preamplifier, such as, the source follower per detector (SFD) circuit can achieve very high sensitivity, but the detector bias voltage changes during integration. On the other hand, non-integration type amplifier, such as the normal transimpedance amplifier (TIA) does not change the bias voltage, but it could not achieve very high sensitivity. The buffered direct injection (BDI) [e.g., 14], the shared buffered direct injection (SBDI) [e.g., 15], and the capacitive transimpedance amplifier (CTIA) can achieve very high sensitivity without the change of bias voltage. However, the BDI and the SBDI are not adequate in applications with ultra low background environments such as astronomical observations, because the transconductances of the injection transistors of the BDI and the SBDI decreases with photocurrent, and these injection efficiencies to an integration capacitance degrade significantly for the extremely low photocurrent such as Ge:Ga detector. On the contrary, the injection efficiency of the CTIA is more than those of the BDI and the SBDI, because the photocurrent is integrated almost only on the feedback capacitance [14], and thus the CTIA is the only circuit that can achieve very high sensitivity in ultra low background. Therefore, it is the best readout amplifier circuit for the astronomical application.

The specifications required for the cryogenic readout amplifier circuits of the FIS are as follows. (1) The operating temperature range is extended down to 1.8 K, the same as that of the Ge:Ga photoconductor array itself. (2) In case of CTIA, the open-loop gain of the amplifier in the CTIA is more than several hundreds for keeping the responsivity of Ge:Ga detector constant within 1 \% during integration. (3) The total power consumption of 175 CTIAs and 35 multiplexers, all that needed for the FIS, is less than 2 mW for the desired cryogen life of 18 months. It means the power consumption for each CTIA must be less than 10 mW. (4) The noise voltage of the cryogenic readout circuit is less than 2 $\mu {\rm V}$/${\sqrt {\rm Hz}}$ at 1 Hz assuming the 1/f noise spectrum and the feedback capacitance of 10 pF for the photon-noise-limited performance in darkest parts of the sky.

\subsection{Circuit Design }
Fig. 1 shows the block diagram of the cryogenic readout circuits for FIS. Five CTIAs are connected to a multiplexer through low-pass filters. The photocurrent generated in each Ge:Ga photoconductor is accumulated in the feedback capacitor. The output voltage change of the CTIA is proportional to the product of the photocurrent and the integration time, and thus a constant photocurrent results in a constant slope of the output voltage. A p-MOSFET is connected in parallel with the feedback capacitor as a reset switch for discharging the integrated charge. The low pass filter (LPFs) between the CTIA and the MPX has the cut-off frequency at 1060 Hz for SW and at 340 Hz for LW. The feedback capacitance is 7 pF for SW and 10 pF for LW. Those values were determined in order to achieve a large output voltage swing within the photon-noise-limited condition during the actual observation. The design and process of the CTIA arrays and the multiplexer are described by Hirao et al. [8]. The design parameters are summarized in Table I.

Fig. 2 shows the schematic diagram of the cryogenic amplifier used here. It corresponds to the symbols of "amplifier" in Fig. 1. The cryogenic amplifier is configured by a three-stage differential amplifier with a source-follower buffered output. In order to control the offset on the output voltage, the level shift circuit is placed between the first and the second-stage differential amplifier. The cascade circuit of each stage reduces the Miller effect. Two power supplies (V1 and V2) and three external bias voltage supplies (VB1, VB2, and VB3) are necessary for each cryogenic amplifier. Unlike commercial operational amplifiers, the cryogenic amplifier is composed only with resistors and p-MOSFETs, because anomalous current-voltage characteristics (kink, hysteresys) at 2K are not suppressed for n-MOSFET [16]. The detailed design parameters of this amplifier are given in caption of Fig. 2. The performances of this amplifier have presented by Hirao et al. [8]. 

\subsection{Fabrication}
Fig. 3 shows the photograph of wafer and reticle layout. There are 60 reticles in this wafer. In a reticle, the 5-channel and 60-channel CTIA arrays, multiplexer, and other test integrated circuits are included. These integrated circuits were fabricated by a standard BiCMOS process with 0.5 $\mu$m design rule. The wafer used was a standard one. In order to improve the performances of p-MOSFET at below liquid helium temperature, the n-well of the p-MOSFET are surrounded by an n+ layer with an n+ collector contact (See "Fig.3" in Hirao et al. [8]).

Fig. 4a and Fig. 4b show the chip layouts of the 5-channel CTIA array for LW and the 60-channel CTIA array for SW, respectively. The chip sizes of 5-channel and 60-channel CTIA arrays are 2 $\times$ 4 $\times$ 0.45 mm (thickness) and 4 $\times$ 14 $\times$ 0.55 mm (thickness), respectively. As for the LW array, the thickness of the substrate was slightly decreased because of its severe volume constraint. The chip surface was covered with an aluminum layer to prevent the incidence of photoemission from MOSFETs on the photoconductor. This phenomenon was demonstrated in the visual wavelength region by [17], and, even in the far-infrared region, the photoemission might affect the detector current and/or noise. The structure of the 60-channel CTIA array chip is more complicated than that of the 5-channel CTIA array in order to hybridize it directly to the Ge:Ga monolithic photoconductor array. In each unit cell of the 60-channel CTIA array, an input terminal is fabricated for the indium bumping to the Ge:Ga monolithic photoconductor array with 3 $\times$ 20 pixels. The LPF circuit with the cut-off frequency at 1060 Hz is also fabricated after each output on the 60-channel CTIA array chip. Additionally, the 60-channel CTIA array has four sets of the electrode for the three voltage supplies, V1, VB1, and VB2. Each electrode set connects to each 15 CTIAs, and helps to reduce the variations of the performances from place to place in the 60-channel CTIA array by tuning each voltage supplies set independently.

\section{PERFORMANCE MEASUREMENTS}
We have measured the performance of the cryogenic readout circuit made here at 2 K. The configuration is shown in Fig. 5. A 5-channel CTIA array, a MPX chip, and a Ge:Ga photoconductor chip were connected in order to simulate the actual configuration of the cryogenic readout circuit of the LW array. A far-infrared light source was placed near the detector chip. Those parts are cooled down to cryogenic temperatures. In the ambient temperature, a JFET-input operational amplifier was connected as a buffer. An additional feed-through wire was connected from the CTIA output to the outside of the cryostat for monitoring the output signal directly. It should be noted that we also confirmed the 5-channel CTIA chip and the MPX chip to be operational at 4.2 K.    

\subsection{Integration Curves and Power Dissipation}
Fig. 6 shows examples of the output signal when the light source was turned on and off. The supplied voltages are summarized in Table II. The CTIA output is almost linear during integration, and the slope of the output signal clearly corresponds to the incident light. The output voltage is -40 mV at the start of integration, and is -310 mV after saturation. It means the output voltage swing is 270 mV. When the light source was on, the slope of the integration curve was almost constant. The linearity of output signal, which is the change of the slope of output signal during integration, is less than 1 \%. By comparing the linearity to that derived from equation (1), which is the model equation of CTIA, the open-loop gain of the CTIA is determined.
\begin{center}
\begin{equation}                
{\rm V_{out}(t)}=-AV_{\rm det}\left[1-{\rm exp} \left(-\frac{t}{C_{f}R_{d}(1+A)}\right)\right]
\end{equation} 
     \end{center}
where, ${\rm V_{out}}$, t, A, ${\rm V_{det}}$, ${\rm C_{f}}$, and ${\rm R_{d}}$, are output voltage of the CTIA, integration time, the open-loop gain of the CTIA, the bias voltage, the feedback capacitance, the impedance of the Ge:Ga photoconductor, respectively. The slope of the integration curve (equation (2)) is derived from the time differentiation of equation (1). 
\begin{equation} 
 {\rm \frac{dV_{out}(t)}{dt}}={\rm exp} \left(-\frac{t}{C_{f}R_{d}(1+A)}\right)
\end{equation} 
The impedance, Rd, which is derived from the slope of the output signal of Fig. 6a, is 6.5 $\times$ 1011 $\Omega$. Then, the linearity of better than 1\% corresponds to the open loop gain of more than 300. However, it is difficult to derive the open loop gain more accurately from the output signal, because the change of the slope is small due to the very large detector impedance. When the light source was off, the slope was almost zero. The current derived from the slope, corresponding to the sum of the photocurrent and the leak current, is less than 1 fA. The power dissipation of the 5-channel CTIA chip is 46 $\mu$W corresponding to 9.2 $\mu$W per each channel, and that is 10.1 $\mu$W per each multiplexer chip.

\subsection{Noise}
Fig. 7 shows the measured noise spectrum at the CTIA output under the dark condition. It is clear that the 1/f noise component is dominant in the measured frequency range. In the range of higher than several hundred hertz, a more flat noise component becomes dominant. The power spectrum density of the noise is 4 $\mu {\rm V_{rms}}$/${\sqrt {\rm Hz}}$ at 1 Hz. Since the output integration signals of the CTIAs are processed by correlated-double-sampling (CDS), we derived the rms voltage noise from the measured noise spectrum by using CDS. The read noise expected in case of the CDS is given by the following equation. 
 \begin{equation}                
{\rm V_{n. rms}^{2}}=\int_{0}^{\infty}e_{n}(f)^{2}\,
\frac{4{\rm sin}^{2}(\pi f T)}{1+\left(\frac{f}{f_{c}}\right)}\,df
\end{equation}      

where, $\rm e_{n}(f)$, t, and ${\rm f_{c}}$ are a power spectral density (PSD) of the read noise at the frequency, f, an integration time, and a cut-off frequency of the low pass filter, respectively. The cut-off frequency of the low pass filter is 340 Hz, which is the same as that of the LW array. Since the integration times of FIS are 0.17 second for SW and 0.28 second for LW in the present stage [9], the integration time of 0.28 second was selected here. The rms voltage noise of the CTIA, Vn,rms, was 16 mV. This corresponds to 1000 electrons for the feedback capacitance of 10 pF. It is worth noting that kTC noise does not contribute to the read noise in this case, because the CDS eliminates kTC noise by sampling the difference between two successive data after a reset. 
It should be noted that, in the section III, we describe the performance of an individual channel of the 5-channel CTIA chip. Performances as a CTIA array such as crosstalk and variations of individual CTIA-performances for different channels are being tested now. They will be published in a forthcoming paper.

\subsection{Comparison to the Requirements for FIS onboard ASTRO-F}
The measured performance is summarized in Table III. Here, we compare these results with the requirements for FIS onboard ASTRO-F on the five points described in section II-A. (1) The cryogenic readout circuit of the present work can be operated at less than liquid helium temperature. Therefore, the readout circuit can be applied to the LW and SW array detectors of FIS. (2) The open-loop gain of the CTIA is more than 300. It means the change of the bias voltage is less than 1 \% for the Ge:Ga of the SW array and 3 \% for stressed Ge:Ga of the LW array. The change of the bias voltage for stressed Ge:Ga detector might be slightly more than the requirement. However, it is acceptable for the requirement in this case, because the change of the responsivity can be avoided by shortening the reset interval. (3) The average power dissipation of the CTIA is 9.2 $\mu$W, and that of the multiplexer chip is 10.1 $\mu$W. The total power dissipation of the all LW and SW arrays is 1.96 mW. This value satisfies the requirement. (4) The output voltage swing of the CTIA is 270 mV. In case of the feedback capacitor of 10 pF, the well depth (the number of electrons that can be accumulated in the feedback capacitance) is $1.7 \times 10^{7}$ electrons. This result is acceptable for the requirement. (5) The voltage noise of the cryogenic readout circuit is 1/f noise of 4 $\mu {\rm V_{rms}}$/${\sqrt {\rm Hz}}$ at 1 Hz. This result is two times larger than that of the requirement. However, it is not so far from the requirement.

In summary, all results except voltage noise are acceptable for the requirements of the FIS. We discuss about improvement of the performance of signal to noise (S/N) ratio in the next section. 

\section{DISCUSSION}
We have achieved the development of the CTIA electronics that can be operated below the liquid-helium temperature. The CTIA has shown good performances of low power consumption, low voltage noise, and sufficient open-loop gain, which are suitable for high accuracy astronomical observations in the far-infrared region. 

By using this design, we fabricated 60-channel readout preamplifier arrays for the Japanese infrared astronomical satellite, ASTRO-F. This preamplifier array consists of CTIA with a reset switch and a low-pass-filter per each channel. This array was successfully bumped with a Ge:Ga monolithic photoconductor array into a complete far-infrared array detector. Details are described in [11]. 

The present work contributes to the establishment of electronics operated at very low temperatures, even at 2 K. We employed a standard process and commercially available, standard silicon wafers without any special requirements. Therefore, widely used LSI technologies could be directly applied to fabricate larger-scale cryogenic integrated circuits: the present 60ch-CTIA design (Fig. 4b) contains 1,800 transistors per chip. Moreover, having two differential inputs, the cryogenic amplifier used in the CTIA is an ordinary operational amplifier. The present work has advantage in these points, which are important for a broader range of application, for example, those for bolometers and super tunneling junction devices.

\subsection{Comparison to the Previous Work}
Cryogenic readout circuits had so far been developed for other infrared astronomical missions, ISO, SIRTF, and SOFIA (the Stratospheric Observatory For Infrared Astronomy) [21]. The present work has clearly improved in several points for that of ISOPHOT [18], and is comparable with that of MIPS onboard SIRTF [6] and that of AIRES onboard SOFIA [20]. Table IV summarizes the performances of these four types of readout circuits.

The first point is the number of readout circuits in one chip. We have developed two-dimensional preamplifier array that has 3 $\times$ 20 CTIA circuits in one chip. This preamplifier array has the largest format in readout circuits for far-infrared detectors. Using the 60-channel preamplifier array, Fujiwara et al. [11] first succeeded in developing direct hybrid two-dimensional array detectors.

The second point is the amplifier gain of the CTIA. The amplifier gain of CTIA directly relates to the constancy of the responsivity of detector; low amplifier gain may introduce changes in the bias voltage. The first CTIAs for Ge:Ga detectors, used as the readout circuits of ISOPHOT, did not have an enough amplifier gain to keep the detector bias voltage constant, and introduced responsivity changes in each integration operation [18]. Young et al. [6] developed the CTIA of very large amplifier gain, CRC696, and obtained the linearity of better than 1 \%. Erickson et al. [22] and Young et al. [23] improved the CRC696 for optimization for high background condition of SOFIA, and named SBRC190. The amplifier gain of SBRC 190 is 500. This value is lower than that of CRC696 because it employed the ac-coupling input through 10 pF. We have developed a CTIA whose amplifier gain is more than 300. This value is sufficient for compensating the change of responsivity of Ge:Ga detectors during integration, although the value of amplifier gain may be smaller than that of Young et al. [6] [23] and Erickson et al. [22].

The third point is the noise performance. As shown in Table IV, the read noise in electron number of the present work (1000 ${\rm e^{-}}$) is larger than those of the previous work. However, this is primarily due to the larger feedback capacitance used in the present CTIA. If compared in voltage (converted from the number of electrons with the feedback capacitance), the noise performance of the present work (16 $\mu$V) is better than those of the previous work.

The final point is the dynamic range. This parameter determines the range of brightness to be observed per unit integration time. The dynamic range is generally defined as the ratio of the well depth to the read noise (e.g., [14]). Since both the well depth and the read noise change linearly with the feedback capacitance when the noise voltage is independent of the capacitance (it is the case if current noise is negligible), the dynamic range can be directly compared although the feedback capacitances of these readout circuits are different from each other. The dynamic ranges are listed in Table 4. The value of the present work is the second largest. Consequently, the readout circuit will have larger dynamic range as well as better read noise performance than those of MIPS and ISOPHOT.

\subsection{Improvement of the Noise Performance}
In the present work, the noise performance does not fully satisfy the requirement of FIS. In order to improve the signal-to-noise (S/N) ratio, we changed the feedback capacitance into smaller values (3 pF). By changing the feedback capacitance from 10 pF to 3 pF, kTC noise, which is inversely proportional to the square root of feedback capacitance, increases from 1.7 $\mu$V to 3.0 $\mu$V in this case. However, as suggested in section III-B, kTC noise does not contribute to the S/N ratio. A preliminary result shows that the noise spectrum is comparable with that in Fig.7, and thus the current noise is negligible. As a result of this improvement, three times higher S/N ratio is expected, and then, the detection limit will be acceptable for the FIS onboard ASTRO-F. 

\section*{Acknowledgement}
We thank Drs. M. Akiba and M. Fujiwara for their important suggestions. We also thank Dr. D. K. Ojha for his critical reading of this paper. We are thankful to all ASTRO-F team members for many discussions and supports. This work was financially supported by a Grant-in-Aid for Scientific Research (No. 10354002) and a Feasibility Study of the Research for the Future program of the Japan Society for the Promotion of Science (JSPS).

\newpage
\begin{table}
\begin{center}
\caption{DESIGN PARAMETERS}
\vspace{3mm}
\begin{tabular}{|c|c|c|}\hline
Detector Array& Ge:Ga 5$\times$20 array & \hspace{5mm}Stressed Ge:Ga 5$\times$15 array\\\hline
Format&60ch CTIA $\times$ 2& 5ch CTIA $\times$ 15 \\
&(3 $\times$ 20 and 2 $\times$ 20)&\\\hline
Process  &\multicolumn{2}{c|}{0.5$\mu$m BiCMOS process}\\\hline
Circuit type &\multicolumn{2}{c|}{CTIA}\\\hline
Feedback capaciancer&7 pF & 10 pF\\\hline
Operating Temperature&2.5 K&1.8 K\\\hline
\end{tabular}
\end{center}
\vspace{15cm}
\end{table}

\newpage
\begin{table}
\begin{center}
\caption{VOLTAGES SUPPLIED TO 5 CHANNEL CTIAS AND MULTIPLEXER}
\vspace{3mm}
\begin{tabular}{|c|c|c|}\hline
Names &Supply Values&Waveform\\\hline\hline
VB1&308 mV&DC\\\hline
VB2&450 mV&DC\\\hline
VB3&-3.1 V&DC\\\hline
V1&1.8 V&DC\\\hline
V2&-1.8 V&DC\\\hline
Vin&90 mV&DC\\\hline
VDD&1.8 V&DC\\\hline
VSS&-0.5 V&DC\\\hline
Reset&&pulse\\
&0.5 V(OFF)/-2.8 V(ON)&20 sec (OFF)/3 msec (OFF))\\\hline
MPX-switch&&pulse\\
&0.5 V(OFF/-2.8 V(ON)&20 sec (OFF)/3 msec (OFF))\\\hline
\end{tabular}
\end{center}
\vspace{10cm}
\end{table}

\newpage
\begin{table}
\begin{center}
\caption{PERFORMANCE OF THE PRESENT WORK}
\vspace{3mm}
\begin{tabular}{|c|c|}\hline
Operating Temperature&1.8 K--4.2 K\\\hline
Open loop gain of CTIA& more than 300\\\hline
Average Power Dissipation of CTIA &9.2 $\mu$W\\\hline
Dynamic Range&270 mV\\\hline
Voltage Noise &4 $\mu \rm{V_{rms}}/\sqrt{\rm{Hz}}@1\rm{Hz}$\\\hline
   \end{tabular}
 \end{center}
 \vspace{15cm}
 \end{table}

\newpage
\begin{table}
\begin{center}
\caption{COMPARISON OF CRYOGENIC READOUT CIRCUITS}
\vspace{3mm}
\begin{tabular}{|c|c|c|c|c|}\hline
MISSION/INSTRUMENT	&	ISO/ISOPHOT  	&	SIRTF/MIPS 	&	SOFIA/AIRES	&	ASTRO-F/FIS	\\
      (Circuit)	&	(CRE11)	&	(CRC696)	&	(SBRC190)	&	  (FA5, FA60)  	\\\hline
 Readout Type	&	CTIA	&	  CTIA	&	  CTIA	&	CTIA \\\hline	
Amplifier Input Type	&	Single-Input	&	Single-Input
 &	Single-Input	&	{\bf Two Differential} 	\\
&&&&{\bf Inputs}\\\hline
Input Coupling	&	AC coupling	&	DC coupling	&	AC coupling	&	{\bf DC Coupling}	\\\hline
Number of Channels	&	11 ch	&	32 ch	&	32 ch	&	{\bf 5 ch / 60 ch}	\\\hline
Feedback Capacitance	&	90 fF	&	36 fF	&	20 fF - 6.8 pF	&	7 pF, 10 pF	\\
(${\rm C_{f}}$)	&		&		&	(Selectable)	&\\\hline
Operation Temperature	&	1.5 - 3.4 K	&	1.5 - 2.0 K	&	1.7 - 2.0 K	&	1.8 - 3.0 K	\\\hline
Integration Time	&	0.92 s	&	8 s	&	0.01 s	&	0.28 s	\\\hline
Amplifier Gain of CTIA	&	56	&	$\gg$1000	&	1.5 $-$ 500	&	$>$ 300	\\\hline
Well Depth	&      $1.1\times 10^{6}$ e-	&$	3.1\times
 10^{5}$ ${\rm e^{-}}$	&	$1.1\times 10^{5}$ - $5.7\times 10^{7}$ ${\rm e^{-}}$ &  $1.1\times 10^{7}$ ${\rm e^{-}}$	\\\hline
Power Dissipation	&	4.5$\sim$8.2 $\mu$W	&	1.3 $\mu$W	&	Unpublished	&	11 $\mu$W	\\\hline
Read Noise	&	260 ${\rm e^{-}}$	&	92 ${\rm e^{-}}$	&	200 - 1000 ${\rm e^{-}}$	&	1000 ${\rm e^{-}}$	\\
(Number of electrons)	&		&		&	(${\rm C_{f}}$=20 fF - 6.8 pF)	&	(${\rm C_{f}}$=10 pF)	\\\hline
Read Noise	&	500 $\mu$V	&	440 $\mu$V	&	24 - 1600 $\mu$V	&	{\bf 16 $\mu$V}	\\
(Voltage)	&		&		&	(${\rm C_{f}}$=6.8 pF - 20 fF)	&		\\\hline
Well Depth/Read Noise	&	4230	&	3370	&	550--57000	&	10000	\\
(=Dynamic Range)&&&&\\\hline
\multicolumn{2}{l}{\hbox to 0pt{\parbox{180mm}{\footnotesize
\vspace{5mm}
These data were taken from [6, 16-18, 21, 22]. The amplifier
 gain of $\gg$1000 for CRC696 was estimated from a typical detector bias of
 50mV [23] and the change of the detector bias of a few mV throughout an
 integration [17]. The amplifier gains of SBRC190 were derived by the ratio of the input capacitance and the feedback capacitances. In this case, the input capacitance is 10 pF, and the feedback capacitances are 20 fF - 6.8 pF. The well depth of ISOPHOT was derived from the dynamic range of 1.9 V and the feedback capacitance of 90 fF. The integration time was derived from the read noise of 260 electrons, the feedback capacitance of 90 fF, and the standard deviation of the integration curve of 0.5 mV/s given in [16]. The values of the read noise in voltage were derived from the ratio of the number of read noise electrons given by [16-18] to the feedback capacitance.}}}
  \end{tabular}
 \end{center}
\end{table}

\begin{figure}
\begin{center}
\FigureFile(145mm,110mm){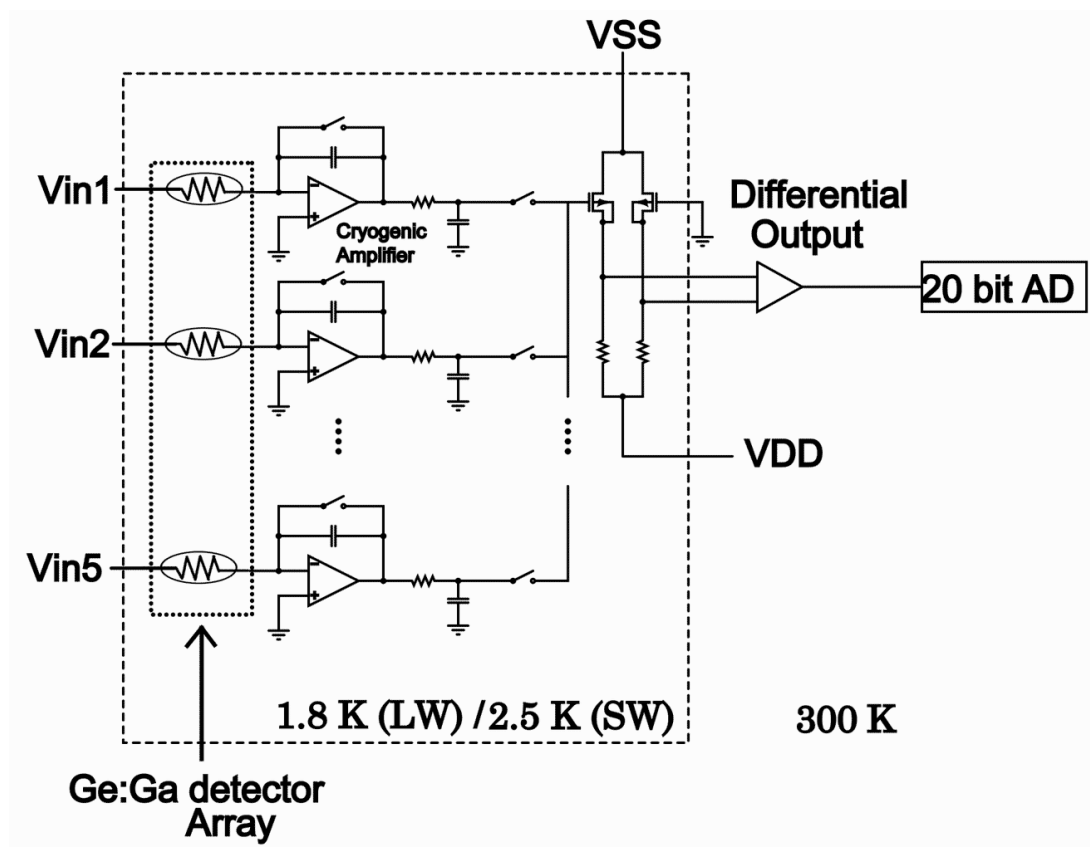}
\caption{Block diagram of the cryogenic readout circuit of FIS onboard ASTRO-F. CTIAs and multiplexers (inside the square) are operated at 1.8 K (LW) and at 2.5 K (SW), respectively.}
\end{center}
\vspace{15cm}
\end{figure}

\begin{figure}
\begin{center}
\FigureFile(150mm,90mm){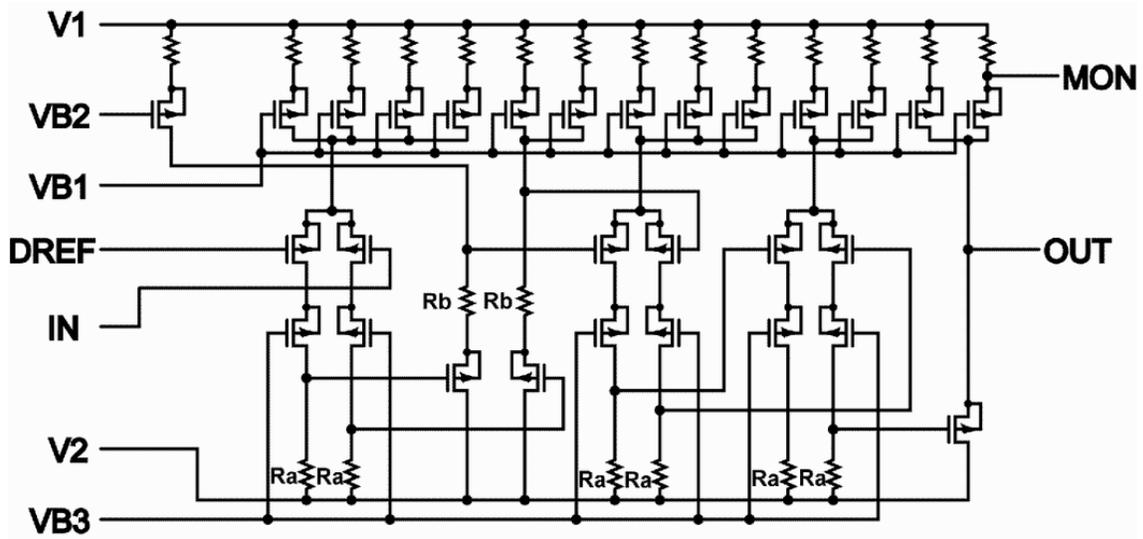}
\caption{ 
Circuit design of the cryogenic amplifier taken from [8]. The gate size of each p-MOSFET is 20.6 $\mu$m in gate width by 5 $\mu$m in gate length. The resistances of these polysilicon resistors at 2 K are Ra=1 M$\Omega$, Rb=200 k$\Omega$, and the other resistors =100 k$\Omega$.
}
\end{center}
\vspace{15cm}
\end{figure}

\begin{figure}
\begin{center}
\FigureFile(133.46mm,132.98mm){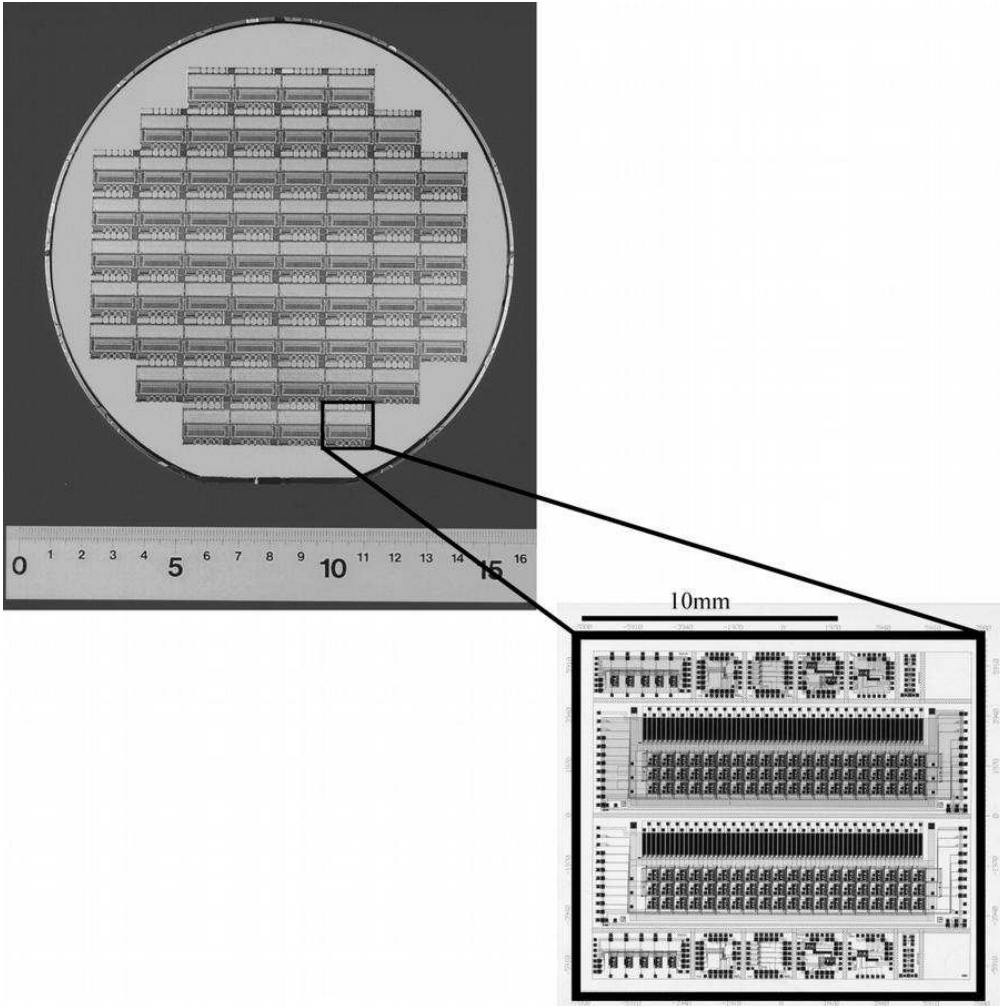}
\caption{ 
A wafer photograph (upper left) and reticle layout (lower right). In a reticle, there are two sets of integrated circuits. The reticle size is 14 mm $\times$14 mm. The upper half part is covered with an aluminum layer to block the emission from the p-MOSFETs (See section II-C). 
}
\end{center}
\end{figure}

\begin{figure}
   \begin{center}
\FigureFile(88.39mm,162.19mm){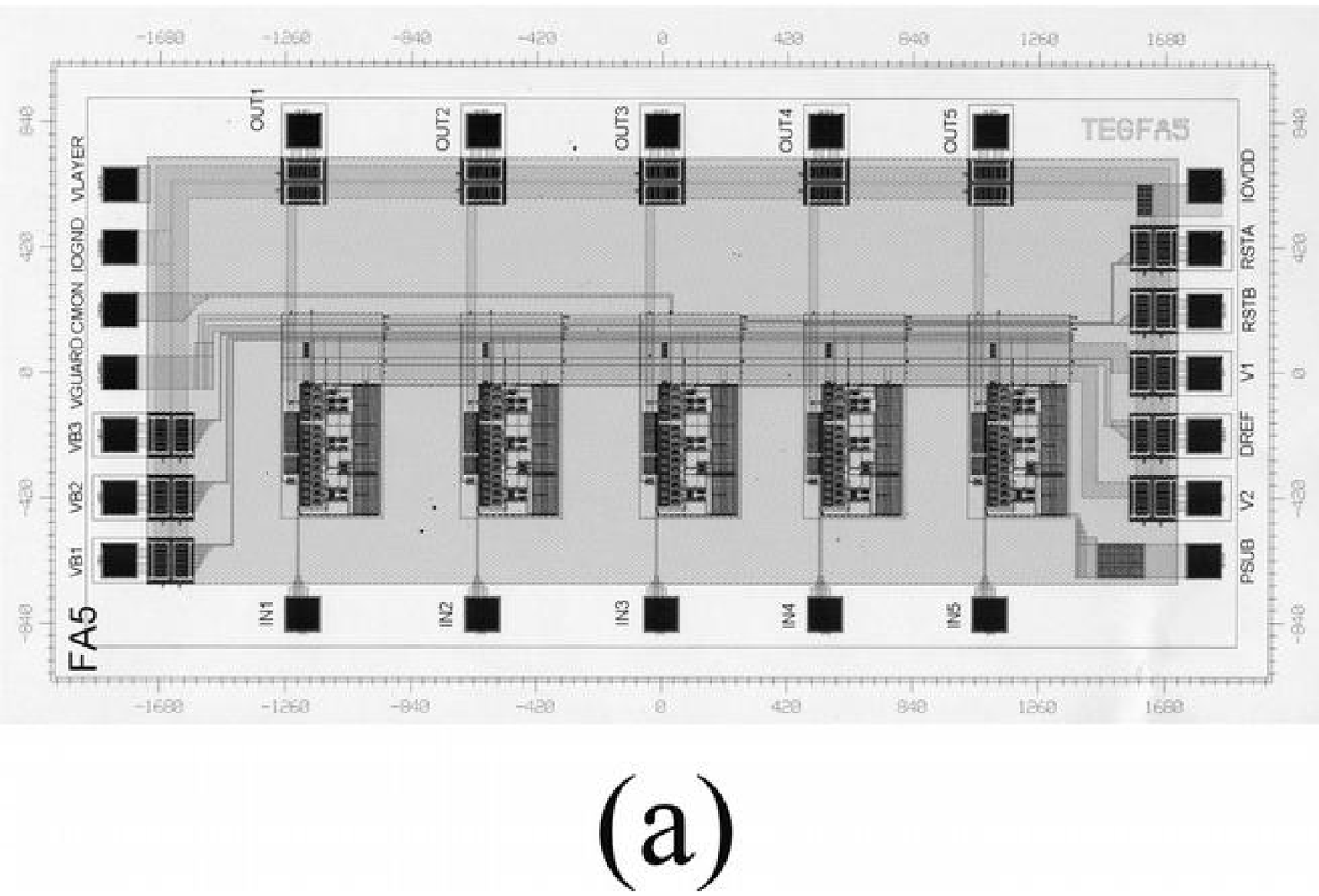}
    \end{center}

   \begin{center}
\FigureFile(102.11mm,142.61mm){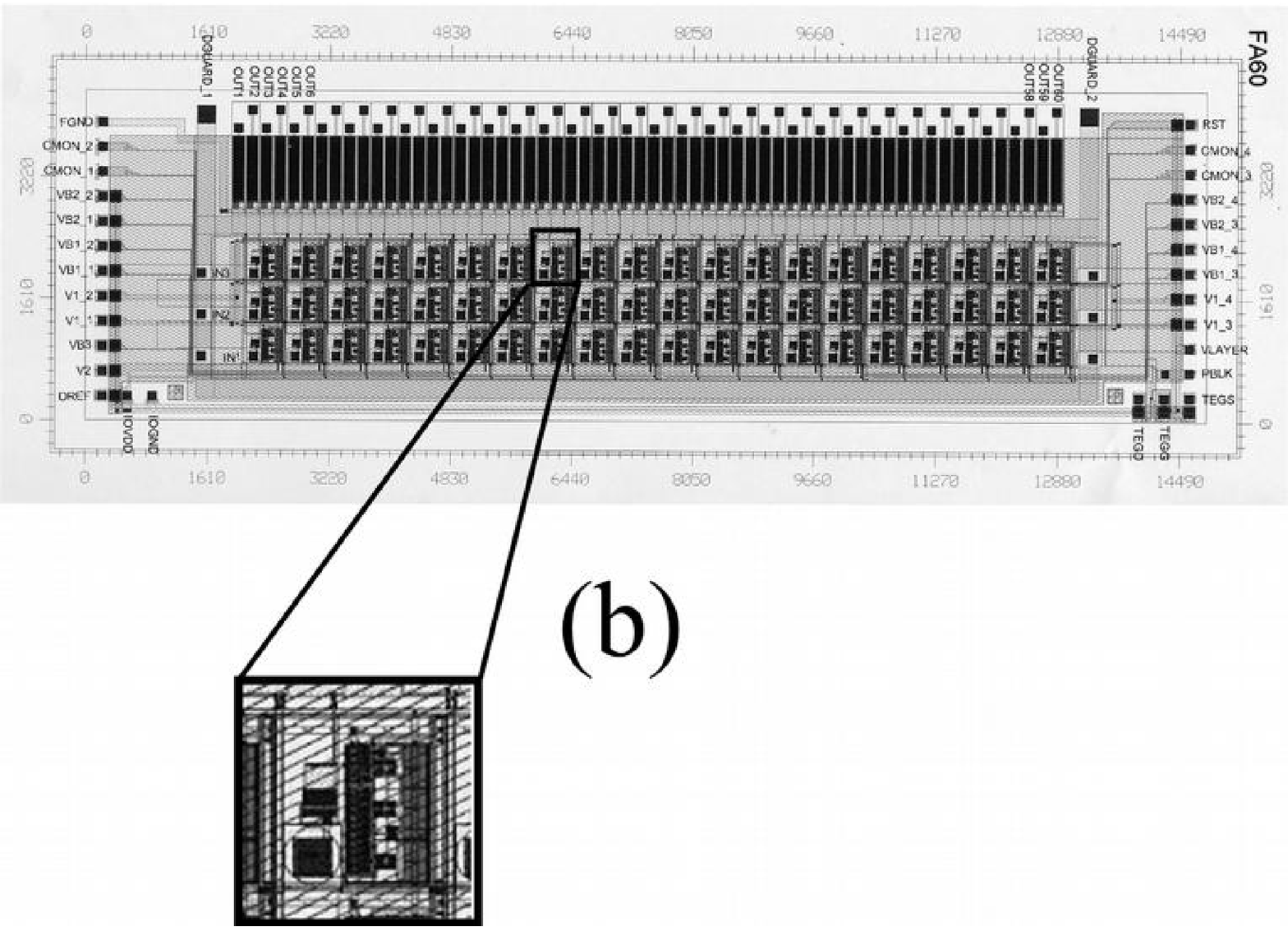}
   \end{center}
\caption{
Layouts of (a) 5-channel CTIA chip and (b) 60-channel CTIA chip. The
 sizes are 2 mm $\times$ mm and 4 mm $\times$ 14 mm, respectively. }
\end{figure}

\begin{figure}
\begin{center}
\FigureFile(135mm,90mm){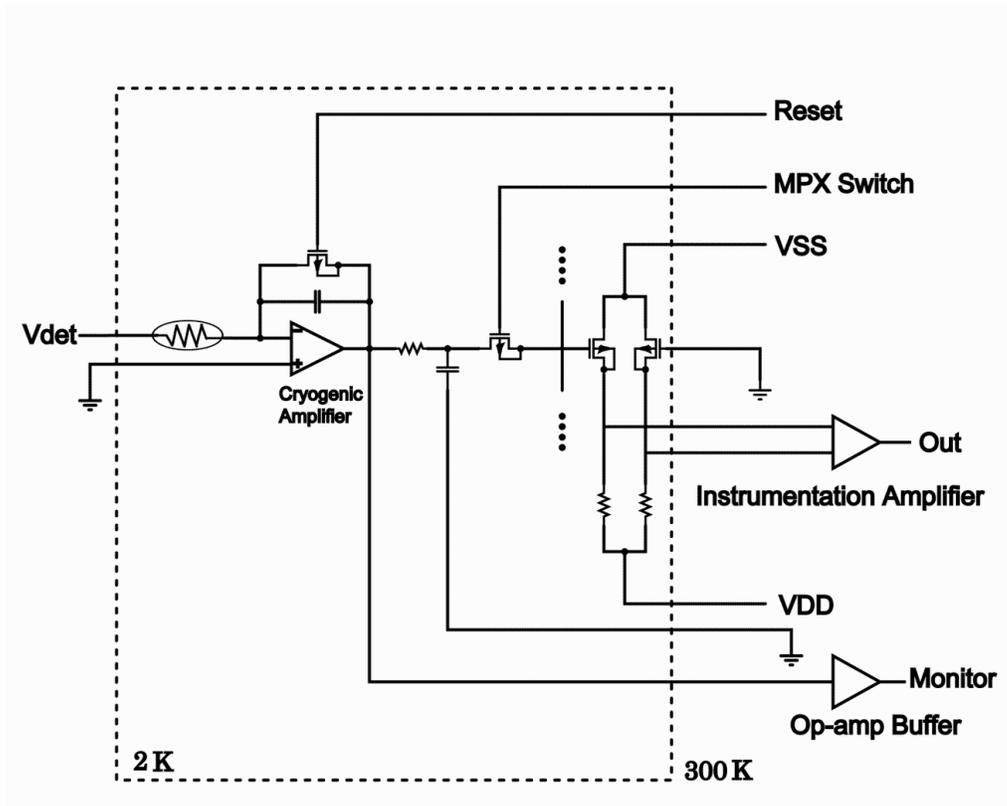}
\end{center}
\caption{
Configuration for the performance measurement of a 5-channel CTIA chip and a multiplexer chip. This is equivalent to that of the LW array of the FIS except for the monitor line from the CTIA output. A Ge:Ga photoconductor chip (0.5 mm cube) was attached to one of the CTIA input. 
}
\end{figure}

\begin{figure}

   \begin{center}
	\FigureFile(80mm,60mm){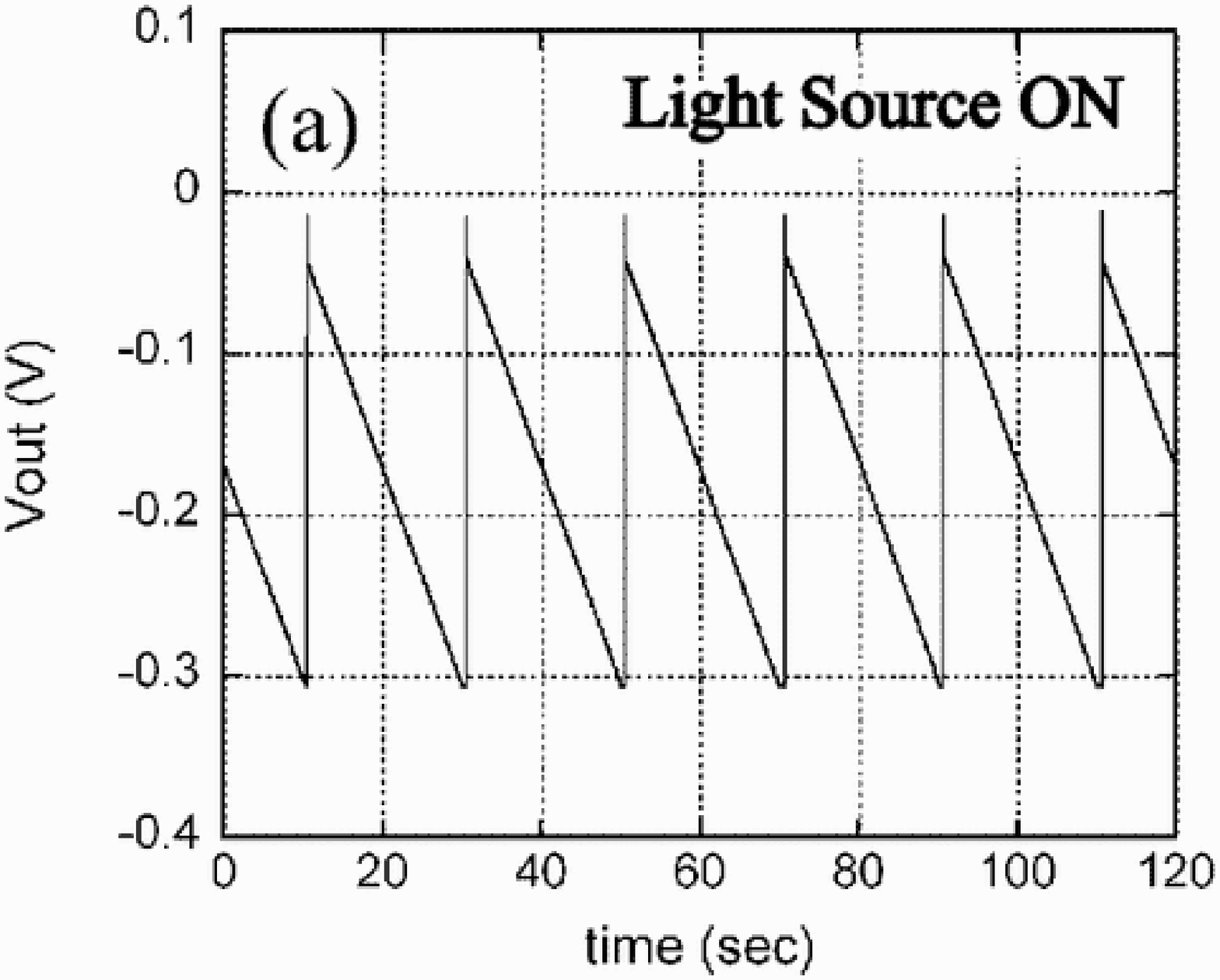}
   \end{center}
   \begin{center}
	\FigureFile(80mm,60mm){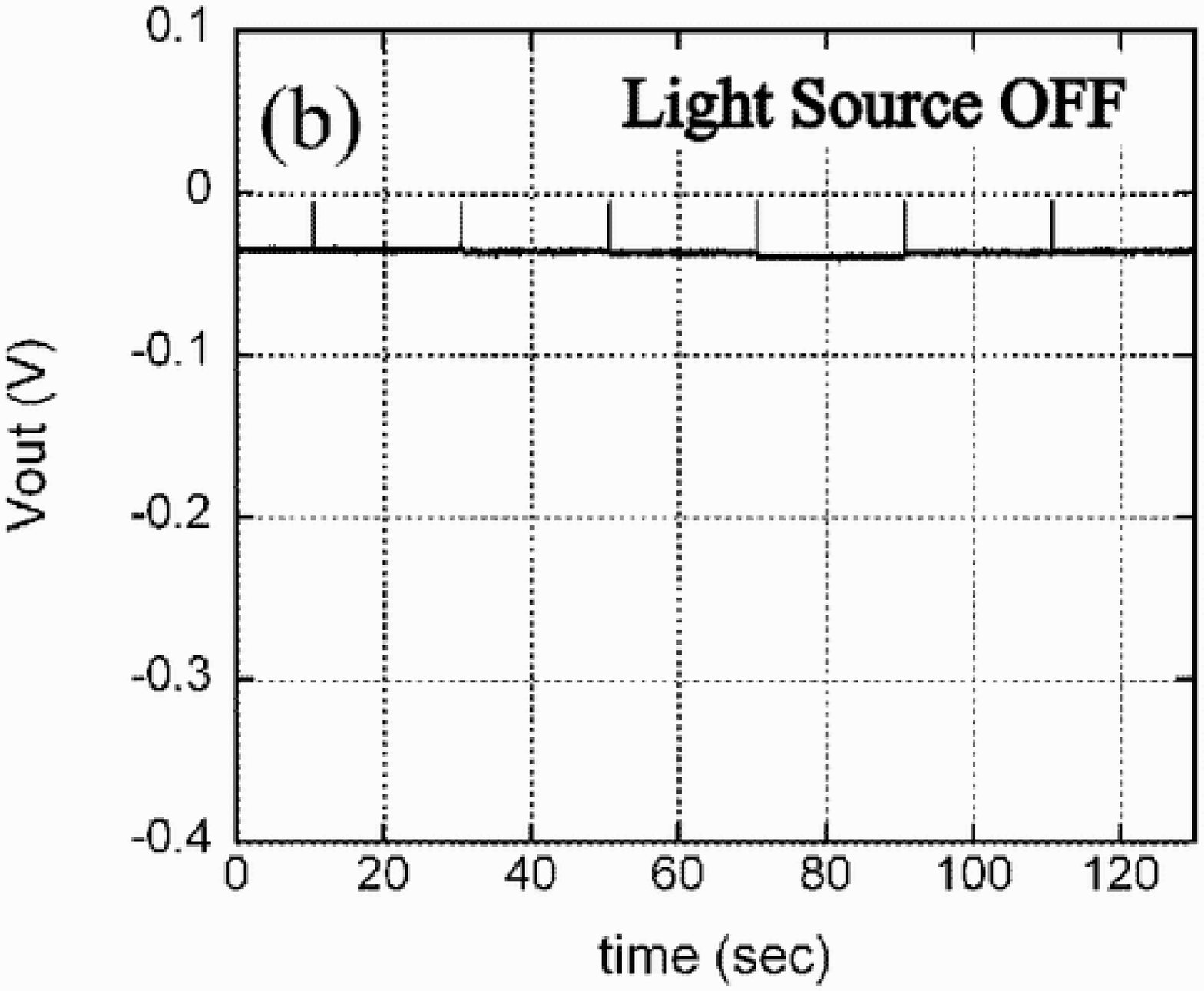}
   \end{center}

\caption{
Output signal during integration with periodical resets; (a) the light source is on, and (b) the light source is off.}
\end{figure}

\begin{figure}
\begin{center}
\FigureFile(160mm,123mm){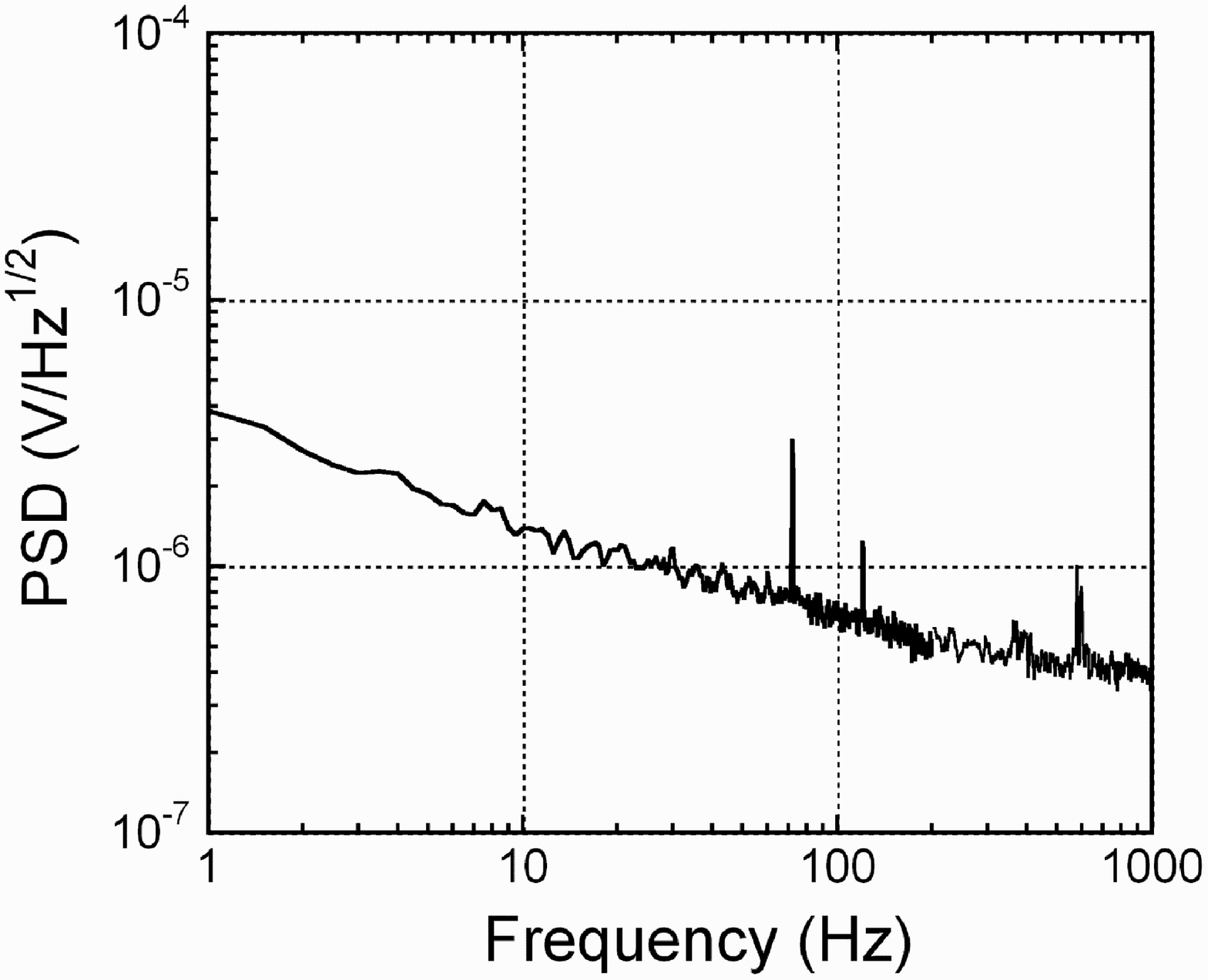}
\end{center}
\caption{ 
Output noise spectrum of the CTIA with a dummy resistor instead of the detector. The sharp peaks at 70, 120, 600 Hz may be due to the electrical interferences in the laboratory environment.
}
\end{figure}


\begin{thebibliography}{}
\bibitem[lemke(2001)]{0}
[1] R. M. Glidden, S. C. Lizotte, J. S. Cable, L. W. Mason, and C. Cao, "Optimization of Cryogenic CMOS Processes for Sub-10K Applications," Proc. SPIE, vol. 1684, pp. 2-39, 1992.
\bibitem[lemke(2001)]{1}
[2] D. Lemke, U. Klaas, J. Abolins, P. Abraham, J. Acosta-Pulido, S. Bogun, H. Castaneda, L. Cornwall, L.  Drury, C. Gabriel, F. Garzon, H. P. Gemuend, U. Groezinger, H. Gruen, M. Haas, C. Hajduk, G. Hall, I. Heinrichsen, U. Herbstmeier, G. Hirth, R. Joseph, U. Kinkel, S. Kirches, C. Koempe, W. Kraetschmer, E. Kreysa, H. Krueger,  M. Kunkel, R. Laureijs, P. Luetzow-Wentzky, K. Mattila, T. Mueller, T. Pacher, G. Pelz, E. Popow, I. Rasmussen, J. Rodriguez Espinosa, P. Richards, S. Russell, H. Schnopper, J. Schubert, B. Schulz, C. Telesco, C. Tilgner, R. Tuffs, H. Voelk, H. Walker, M. Wells, and J. Wolf, "ISOPHOT - capabilities and performance," Astronomy \& Astrophysics, vol.315, pp. L64-L70, 1996.
\bibitem[Author(2001)]{2}
[3] M. F. Kessler, J. A. Steinz, M. E. Anderegg, J. Clavel, G. Drechsel, P. Estaria, J. Faelker, J. R. Riedinger, A. Robson, B. G. Taylor, and F. S. Ximenez, "The Infrared Space Observatory (ISO) mission, " Astronomy \& Astrophysics, Vol. 315, pp. L27-L31, 1996.
\bibitem[Author(2001)]{3}
[4] B. Dierickx, E. Simoen, J. Vermeiren, S. Cos, C. Claeys, and G. Declerck, "Optimization of CMOS Technology and Design for Deep Cryogenic Analog Circuits," in Proc. ESA Electronic Components Conf., vol. ESA-SP 313, pp. 43-47, 1991.
\bibitem[Author(2001)]{4}
[5] J. L. Fanson., G. G. Fazio., J. R. Houck., T. Kelly., G. H. Rieke., D. J. Tenerelli., M. Whitten, "Space Infrared Telescope Facility (SIRTF)," Proc. SPIE, Vol. 3356, pp. 478-491, 1998.
\bibitem[Author(2001)]{5}
[6] E. T. Young, J. T. Davis, C. L. Thompson, G. H. Rieke, G. Rivlis, R. Schnurr, J. Cadien, and L. Davidson, "Far-infrared imaging array for SIRTF," Proc. SPIE, vol. 3354, pp. 57-65, 1998.
\bibitem[Author(2001)]{6}
[7] E.T. Young, G. H. Rieke, H. Dang, I. Barg, and C. L. Thompson, "Test results for the SIRTF far-infrared array module," Proc. SPIE, vol. 2475, pp. 435-440, 1995.
\bibitem[Author(2001)]{7}
[8] T. Hirao, Y. Hibi, M. Kawada, H. Nagata, H. Shibai, T. Watabe, M. Noda, and T. Nakagawa, "CRYOGENIC READOUT ELECTRONICS WITH SILICON P-MOSFETS FOR THE INFRARED ASTRONOMICAL SATELLITE, ASTRO-F," Advances in Space Research, vol. 30, pp. 2117-2122, 2002. 
\bibitem[Author(2001)]{8}
[9] H. Takahashi, H.  Shibai, M. Kawada, T. Hirao, T. Watabe, Y. Tsuduku, H. Nagata, H. Utsuno, Y. Hibi, S. Hirooka, T. Nakagawa, H. Kaneda, S. Matsuura, T. Kii, S. Makiuti, Y. Okamura, Y. Doi, H. Matsuo, N. Hiromoto, M. Fujiwara, and M. Noda, "FIS: far-infrared surveyor on board ASTRO-F (IRIS)," Proc. SPIE, vol. 4013, pp. 47-58, 2000. 
\bibitem[Author(2001)]{9}
[10] H. Shibai, "ASTRO-F: THE INFRARED IMAGING SURVEYOR (IRIS) MISSION," Advances in Space Research, vol. 30, pp. 2089-2097, 2002.   
\bibitem[Author(2001)]{10}
[11] M. Fujiwara, M. Sasaki, T. Hirao, M. Kawada, H. Shibai, S. Matsuura, H. Kaneda, M. Patrashin, and T. Nakagawa, "Development of Ge:Ga far-infrared photoconductor direct hybrid 2D array," Applied Optics-LP, vol. 42, Issue 12, pp.2166-2173, April 2003. 
\bibitem[Author(2001)]{11}
[12] S. Matsuura, Y. Isozaki, M. Shirahata, H. Kaneda T. Nakagawa, M. Patrashin, M. Fujiwara, T. Hirao, M. Kawada, H. Shibai, and T. Watabe, "Monolithic Ge:Ga two-dimensional array detector for FIS instrument on ASTRO-F" Proc. SPIE, vol. 4850, pp. 902-909, 2003
\bibitem[Author(2001)]{12}
[13] Y. Doi, S. Hirooka, A. Sato, M. Kawada, H. Shibai, Y. Okamura, S. Makiuti, T. Nakagawa, N. Hiromoto, and M. Fujiwara, "LARGE-FORMAT AND COMPACT STRESSED GE:GA ARRAY FOR THE ASTRO-F(IRIS) MISSION," Advances in Space Research, vol. 30, pp. 2099-2104, 2002. 
\bibitem[Author(2001)]{13}
[14] E. R. Fossum, and B. Pain, "Infrared Readout Electronics for Space Science Sensors: State of the Art and Future Directions," Proc. SPIE, vol. 2020, pp. 262-285, 1993.
\bibitem[Author(2001)]{14}
[15] C. C. Hsieh, and C. Y. Wu, "Focal-Plane-Arrays and CMOS Readout Techniques of Infrared Imaging Systems" IEEE Trans. Circuits and Systems for Video Technology, vol. 7, No. 4, pp. 594-605, August. 1997.
\bibitem[Author(2001)]{15}
[16] M. Noda, H. Shibai, T. Watabe, T. Hirao, H. Yoda, H. Nagata, T. Nakagawa, and M. Kawada, "Evaluation of charge integrating amplifiers with silicon MOSFETs for cryogenic readout," Proc. SPIE, vol. 3354, pp. 247-252, 1998. 
\bibitem[Author(2001)]{16}
[17] S. Tam, and C. Hu, "Hot-Electron-Induced Photon and Photocarrier Generation in Silicon MOSFET's," IEEE Trans. Electron Devices, vol. ED-31, pp. 1264-1273, Sep. 1984.
\bibitem[Author(2001)]{17}
[18] J. Wolf, U. Grozinger, and D. Lemke, "Cryogenic Readout Electronics for the Infrared Space Observatory ISO," in Third Symposiumon Low Temperature Electronicsand High Temperature Super-conductivity, 1995, pp. 355-368  
\bibitem[Author(2001)]{18}
[19] E. T. Young, G. H. Rieke, J. D. Cadiena, H. A. Dolea, C. W. Englebrachta, K. D. Gordona, G. B. Heimb, D. M. Kellya, and J. A. Stansberrya, "Ground test characterization of the Muliband Imaging Photometer for SIRTF (MIPS)" Proc. SPIE, vol. 4850, pp. 98-107, 2003
\bibitem[Author(2001)]{19}
[20] J. Farhoomand, D. Sisson, L. Yuen, and D. Hoang, "Characterization of SBRC-190: a multi-gain, cryogenic readout multiplexer for IR detector arrays" Proc. SPIE, Vol. 4857, pp. 144-154, 2003
\bibitem[Author(2001)]{20}
[21] E. E. Becklin, S. C. Casey, J. A. Davidson, and M. L. Savage, "Stratospheric Observatory for Infrared Astronomy (SOFIA)" Proc. of SPIE, Vol. 3354, pp. 922-929, 1998.
\bibitem[Author(2001)]{21}
[22] E. F. Erickson, E. T. Young, J. Wolf, J. F. Asbrock, and N. A. Lum, "SBRC-190: A Cryogenic Multiplexer for Moderate-Background FIR Astronomy," presented at Far-IR, Sub-mm \& mm Detector Technology Workshop, Monterey, CA, 2002.  
\bibitem[Author(2001)]{22}
[23] E. Young, G. Rieke, and J. Davis, "Advanced Photoconductor Arrays," presented at Far-IR, Sub-mm \& mm Detector Technology Workshop, Monterey, CA, 2002.  
\bibitem[Author(2001)]{23}
[24] R. Schnurr, C. L. Thompson, J. T. Davis, J. W. Beeman, J. Cadien. E. T. Young, E. E Haller, and G. H. Rieke, "Design of the stressed Ge:Ga far-infrared array for SIRTF Authors", Proc. of SPIE, vol. 3354, pp. 322-331, 1998.
\bibitem[Author(2001)]{24}
[25] E. T. Young, "Progress on readout electronics for far-infrared arrays," Proc. SPIE, vol. 2226, pp. 21-28, 1994.
\end{thebibliography}
\end{document}